\documentclass[pre,showkeys,showpacs,english,aps,twocolumn,a4paper,floatfix]{revtex4}
\usepackage[T1]{fontenc}
\usepackage{color}
\usepackage{amsmath}
\usepackage{amssymb}
\usepackage[latin1]{inputenc}
\usepackage{graphicx}
\usepackage{bm}
\usepackage{epsfig}
\providecommand{\keywords}[1]{\textbf{\textit{Index terms---}} #1}

\begin{document}

\title{Role of length-polydispersity on the phase behavior of freely-rotating hard-rectangle fluid} 

\author{Ariel D\'{\i}az-De Armas}\email{ardiaza@math.uc3m.es}
\author{Yuri Mart\'{\i}nez-Rat\'on} \email{yuri@math.uc3m.es}
\affiliation{Grupo Interdisciplinar de Sistemas Complejos (GISC), Departamento de Matem\'aticas, Escuela 
Polit\'ecnica Superior, Universidad Carlos III de Madrid, Avenida de la Universidad 30, 28911-Legan\'es, Madrid,
Spain.}

\begin{abstract}
We used the Density Functional formalism, in particular the Scaled Particle Theory, applied to a 
length-polydisperse hard-rectangular fluid to study its phase behavior as a function of the mean  
particle aspect ratio ($\kappa_0$) and polydispersity ($\Delta_0$). The numerical solutions of the 
coexistence equations were calculated  
by transforming the original problem with infinite degrees of freedoms to a finite 
set of equations for the amplitudes of the Fourier expansion of the moments of the density profiles. 
We divided the study into two parts: The first one is devoted to the calculation 
of the phase diagrams in the packing fraction ($\eta_0$)- $\kappa_0$ plane for a fixed $\Delta_0$ and 
selecting parent distribution functions with exponential (the Schulz distribution) 
or Gaussian decays. In the second part we study the phase behavior in the $\eta_0$-$\Delta_0$ plane 
for fixed $\kappa_0$ while  $\Delta_0$ is changed. We characterize in detail the orientational 
ordering of particles and the fractionation of different species between the coexisting phases. 
Also we  study the character (second vs. first order) of the Isotropic-Nematic 
phase transition as a function of polydispersity. We particularly focused  
on the stability of the Tetratic phase as a function of $\kappa_0$ and $\Delta_0$. 
The Isotropic-Nematic transition becomes strongly of first order  
when polydispersity is increased: the coexisting gap widens and the location of the tricritical point moves to higher values of $\kappa_0$ while the Tetratic phase is slightly destabilized with 
respect to the Nematic one. The results obtained here can be tested in experiments on shaken 
monolayers of granular rods.  
\end{abstract}

\date{\today}



\maketitle

\section{Introduction}

The size polydispersity is an important feature present in experiments conducted on colloidal suspensions of anisotropic, rod or plate-like, particles 
\cite{Kooij1,Kooij2,Petukhov,Woolston,Vroege,Pol,Verhoeff,Sun,Byelov,Leferik}. 
The process of synthesis of mineral particles can not avoid the presence of polydispersity in  
sizes and only a sequence of fractionation steps can reduce it considerably. 
In experimental situations the number of species with different sizes is so large that 
we can consider a continuous size polydispersity as a reasonable approximation. 
However, the inclusion of a continuous size-polydispersity  
in theoretical models considerably complicates its numerical implementation. 
This constitutes the reason why 
the moments method was developed as a powerful theoretical tool to approximately solve the 
equations resulting from the 
two-phase coexistence calculations \cite{Clarke,Sollich,Speranza1,Speranza2}. The polydispersity is present not only in colloids  
but also in biological \cite{Fraden} and granular \cite{Nguyen} systems. 
The polymerization of F-actine filaments confined in quasi 
two-dimensional geometries constitutes a particular realization of a two-dimensional liquid-crystal 
system  made of polydisperse flexible rods \cite{Viamontes}.  

Several experimental \cite{Kooij1,Kooij2,Petukhov,Woolston,Vroege,Pol,Verhoeff,Sun,Byelov,Leferik} 
and theoretical \cite{Sluckin,Bates1,Bates2,M-R1,M-R2,Wensink,Speranza3,Richter,Green2,Pol2,M-R3,Green1,M-R4,M-R5, 
Escobedo2} 
works have studied the effect of polydispersity in particle sizes 
on the phase behavior and orientational ordering properties 
of colloidal suspensions made of anisotropic particles. 
Some general trends found in these studies concerning to the isotropic (I)-nematic (N) 
phase transition can be 
summarized as follows: (i) the polydispersity dramatically widens the region of I-N 
two-phase coexistence, i.e. the density gap between the cloud and shadow points is considerably 
enlarged making this transition strongly of first order. (ii) There exists strong fractionation 
in sizes between coexisting phases from which small and large particles usually populate the I 
and N phases respectively. The above mentioned  
experimental and theoretical works were conducted on three-dimensional systems, however the 
effect of polydispersity on two-dimensional liquid-crystals has been scarcely explored 
\cite{Ioffe,Tavares,Lopez,Almarza,Constantin}. 
An experimental realization of a quasi 2D liquid-crystal system, apart from that already 
pointed out before, can be obtained by vertically shaking a monolayer of granular rods. 
Recent experiments on these monolayers showed the presence of liquid-crystal 
textures as stationary states \cite{Narayan1,Galanis1,Galanis2,Heras}. 

Theoretical studies on 2D monodisperse hard rods have shown that  
the transition from the high density N phase (without long range orientational order) 
to the low-density I phase is continuous, via a Kosterlitz-Thouless disclination 
unbinding type mechanism \cite{Kosterlitz}, rather than first order \cite{Bates3,Lagomarsino}. 
However, recent studies proved that 
when particle interactions are of a certain type the I-N transition becomes of first order 
in 2D \cite{Aernout,Vink}. 
A recent controversy resulted on the universality class (Ising vs. that corresponding to $q=1$ 
Pott s-type models) of the I-N transition of hard Zwanzig self-assembled rods on a lattice \cite{Lopez,Almarza}: 
A mean-field theory shows that these polydisperse self-assembled rods exhibit a 
first order I-N transition while MC results discard this fact \cite{Linares}. 
Finally, a recent experiment on two-dimensional polydisperse hard rods 
(where magnetic nanorods were strongly confined between layers of 
a lamellar phase) have shown a first order I-N transition \cite{Constantin}. 

The particular shape of 
two-dimensional rods is also determinant to stabilize the N or tetratic (T) phases at low aspect ratios $\kappa$.  
In the latter the angular distribution function is invariant under $\pi/2$ rotations. Hard ellipses (HE)
\cite{Cuesta1,Cuesta2,Xu,Bautista} and hard discorectangles (HDR) \cite{Bates3} only show 
the usual I-N transition followed by transitions to 
the plastic or orientationally ordered crystals. Hard rectangles (HR), 
on the other hand, also show fluid or crystalline phases with tetratic symmetry at low $\kappa_0$
\cite{Schlacken,Frenkel,M-R6,Donev,M-R7,Triplett}. 
The T phase of HR was theoretically predicted by mean-field DF studies \cite{Schlacken,M-R6}. 
MC simulations on hard squares (HS) showed the presence of strong tetratic correlations 
of quasi-long-range order \cite{Frenkel} and simulations on HR of $\kappa_0=2$ showed a 
liquid with the same T correlations and no N order \cite{Donev}. 
The solid phase in the latter system was observed  with a nonperiodic T ordering and 
having the structure of a random tiling of the square lattice with dimers of HR randomly 
oriented \cite{Donev}. 
Experiments on a monolayer of disks standing on edges observed the conventional Kosterlitz-Thouless 
transition from I to N with almost smectic behavior at high density. But on the isotropic side 
of I-N transition 
an unusual regime of short-range T correlations dominates over N ones \cite{Chaikin}. 
Recent experiments on monolayers of hard microscale square platelets showed a phase transition 
between an hexagonal rotator 
crystal and a rhombic crystal as the packing fraction is increased \cite{Zhao}. 
The absence of T ordering in this system was further explained resorting to the roundness 
of the square corners \cite{Escobedo}. 
If the roundness is sufficiently small the particles behave like perfect squares and the T phase is 
recovered \cite{Escobedo}.    
Shaken monolayers of granular cubes \cite{Narayan2} and cylinders \cite{Narayan1,Heras,G-P} 
exhibit stationary textures with strong T ordering even for aspect ratios of cylinders as large as 7. 

All these studies show the profound effect that particle shapes and pair interactions 
have on the symmetry of the orientationally ordered phases. Even the character 
(continuous vs. first order) can be influenced by them. The mean-field SPT 
for HR predicts a first-order I-N (or T-N) 
transition located between two tricritical points. This is a peculiar feature of HR because the I-N transition for 
HDR and HE following the same theory is always of second order \cite{M-R6}. This difference can be explained 
by the peculiar form of the excluded volume between two HR as it will be discussed in Sec. \ref{SPT}. 

The main purpose of the present work is to study how the polydispersity affects the phase behavior of HR. 
To this purpose we will extend the mean-field SPT formalism from its multicomponent version 
(already described in Ref. \cite{M-R8,Heras2}) 
to the length-polydisperse case. The minor length of rectangles is considered constant 
while the major one is polydisperse  
according to a Schulz-type (with exponential decay) probability density distribution or also 
according to a distribution with a Gaussian decay. We calculated the phase diagrams for a 
fixed $\Delta_0$ while $\kappa_0$ varies, and also for 
certain fixed values of $\kappa_0$ while $\Delta_0$ changes. We  
measured the degrees of orientational ordering of particles 
and their fractionation between the coexisting phases.  
The addition of polydispersity dramatically increases the interval of aspect ratios 
where the I-N is of first order and the I-N tricritical point moves to higher values of $\kappa_0$.  
Also we prove that the stability of the T phase is not severely affected by polydispersity: 
the area of the phase diagram  where the T phase is stable  
has not a severe decrement when the polydispersity is increased from zero to its maximum allowed value.

\section{Model and Theory}
In this section we present the theoretical tools used to study the phase behavior of a 
length-polydisperse HR fluid. In Sec. \ref{model} we formulate the HR model 
and define the family of length-polydisperse probability distributions we have used. 
In Sec. \ref{SPT} we generalize the SPT formalism from the multicomponent HR 
mixture \cite{M-R8} to a continuous polydisperse fluid. Further we explicitly obtain the set 
of equations used in Sec. \ref{coexistence} to calculate the two-phase coexistence  
giving also details on their numerical implementation. A concise summary of the I-N, I-T and T-N 
bifurcation analysis will be given in Sec. \ref{bifurca}. Finally, in Sec. 
\ref{measures}, we define the main variables and functions used to measure the orientational 
ordering and fractionation between coexisting phases.

\subsection{Polydisperse HR model}
\label{model}
Our model consists on a collection of freely-rotating HR. They move and rotate in 2D and 
cannot overlap. We take the minor edge-length of rectangles, $\sigma$, 
to be constant while its mayor length, $L$ (with $\sigma<L<\infty$), defining the main particle axis, 
is considered continuously polydisperse according to a fixed, so called \emph{parent}, probability 
density distribution function: 
\begin{eqnarray}
&&f_0(\kappa)=C_{\nu,q}(\kappa_0)
\left(\frac{\kappa-1}{\kappa_0-1}\right)^{\nu}
\exp\left[-\lambda_{\nu,q}\left(\frac{\kappa-1}{\kappa_0-1}\right)^q\right], \nonumber\\ \label{distri0}\\
&&C_{\nu,q}(\kappa_0)=q\left(\kappa_0-1\right)^{-1}
\frac{\Gamma^{\nu+1}\left[(\nu+2)/q\right]}{\Gamma^{\nu+2}\left[(\nu+1)/q\right]}, \\
&&\lambda^{1/q}_{\nu,q}=\frac{\Gamma\left[(\nu+2)/q\right]}{\Gamma\left[(\nu+1)/q\right]},
\end{eqnarray}
defined in terms of the particle aspect ratio $\kappa\equiv L/\sigma$. This 
function is normalized to unity,  
$\int_1^{\infty}d\kappa f_0(\kappa)=1$, and 
its first moment, $\langle\kappa\rangle\equiv\int_1^{\infty}d\kappa \kappa f_0(\kappa)=\kappa_0$,
 is selected to be equal to $\kappa_0$ (the mean aspect ratio).  
The polydisperse coefficient, or mean square deviation, can be calculated as 
\begin{eqnarray}
&&\Delta\equiv \sqrt{\frac{\langle \kappa^2\rangle-\langle \kappa\rangle^2}{\langle \kappa\rangle^2}}
=\left(1-\frac{1}{\kappa_0}\right)\Delta_0,\label{deviation}\\
&&\Delta_0\equiv \sqrt{\frac{\Gamma\left[(\nu+1)/q\right]\Gamma\left[(\nu+3)/q\right]}
{\Gamma^2\left[(\nu+2)/q\right]}-1},
\end{eqnarray}
where $\Gamma(x)$ is the standard Gamma function. The parameters $\nu$ and $q$, used to 
define the family of functions (\ref{distri0}), control the 
width of $f_0(\kappa)$, with $q$ being the one that dictates the decay of the tail: 
for $q=1$ (corresponding to the Schulz distribution function) we obtain an exponential decay 
while for $q=2$ a Gaussian tail is obtained. In the following we will use $\Delta_0$ 
as a measure of the degree of length-polydispersity.  

\subsection{SPT for length-polydisperse HR}
\label{SPT}
\begin{figure}
\epsfig{file=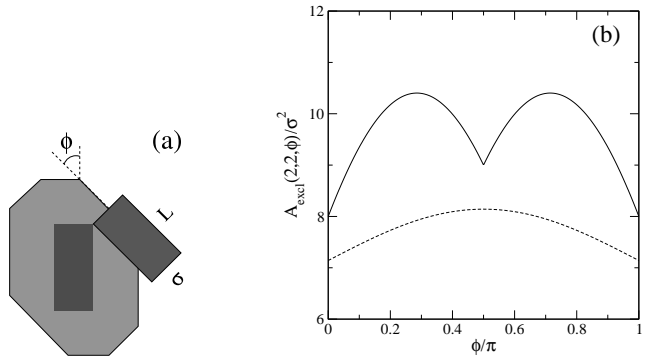,width=0.9in}
\hspace*{1.cm}
\epsfig{file=fig1b.eps,width=1.9in}
\caption{(a): Sketch of the excluded area between two HR  
of aspect ratio $\kappa=L/\sigma=2$ and relative angle $\phi$ between their long axes. 
(b): Excluded areas in units of $\sigma^2$ between two HR (solid) and two 
HDR (dashed) of same aspect ratio $\kappa=2$.}
\label{fig1}
\end{figure}

We use the SPT formalism for a multicomponent HR fluid \cite{M-R8} extended to the 
continuously length-polydisperse limit: $\rho_{\nu}(\phi)\to\rho(\kappa,\phi)$,  
where $\rho_{\nu}(\phi)$ corresponds to the angular density 
profile of species $\nu$ of the multicomponent mixture while $\rho(\kappa,\phi)$ 
denotes the continuously distributed angular density profile of species with aspect ratio 
$\kappa$.  
All sums 
over species in equations used to define the free-energy density in Ref. \cite{M-R8} 
are now substituted  by integrals over the aspect ratio: $\sum_{\nu}\to \int d\kappa$. 
The resulting free-energy 
density in reduced thermal units, 
$\Phi[\rho]\equiv \Phi_{\rm id}[\rho]+\Phi_{\rm ex}[\rho]$, splitted in its ideal and excess 
contributions, $\Phi_{\rm id}[\rho]$ and $\Phi_{\rm ex}[\rho]$ respectively, is a functional of  
$\rho(\kappa,\phi)$:
\begin{eqnarray}
&&\Phi_{\rm id}[\rho]\equiv \beta{\cal F}_{\rm id}[\rho]/A=
\int d\kappa \int d\phi \rho(\kappa,\phi)\left[\log \rho(\kappa,\phi)-1\right]\nonumber\\&&\label{ideal}\\
&&\Phi_{\rm ex}[\rho]\equiv \beta{\cal F}_{\rm id}[\rho]/A
=-\rho_0^{(0)}\ln\left(1-\rho_0^{(1)}\sigma^2\right)\nonumber\\
&&+\frac{S_0\left[\rho\right]}{1-\rho_0^{(1)}\sigma^2},\label{excess}\\
&&S_0\left[\rho\right]\equiv\int d \kappa_1\int d \kappa_2\int d\phi_1\int d\phi_2 
\rho(\kappa_1,\phi_1)\rho(\kappa_2,\phi_2)\nonumber\\
&&\times A_0(\kappa_1,\kappa_2,\phi_1-\phi_2),\\
&&A_0(\kappa_1,\kappa_2,\phi)=\frac{\sigma^2}{2}\left\{(\kappa_1\kappa_2+1)|\sin\phi|\right.\nonumber\\
&&\left.+\left(\kappa_1+\kappa_2\right)|\cos\phi|\right\}.
\end{eqnarray}
In (\ref{ideal}) and (\ref{excess}), $\beta\equiv 1/k_BT$, is the Boltzmann factor, 
${\cal F}_{\rm id,ex}[\rho]$ correspond to the ideal and excess parts of the 
free-energy density functional while $A$ is the total area of the system. We have also defined  
the $i$th moment of the integrated, over $\kappa$ and $\phi$, density profile: 
\begin{eqnarray}
\rho^{(i)}_0\equiv \int_0^{\pi}d\phi \rho^{(i)}(\phi)=
\int_0^{\pi}d\phi\left[\int_1^{\infty} d\kappa\kappa^i\rho(\kappa,\phi)\right],
\label{moments}
\end{eqnarray} 
where $i=1,2$ and the integration with respect to $\phi$ 
was taken between 0 and $\pi$ (instead of 2$\pi$) due to the head-tail 
symmetry of rectangles. Note that these moments can be obtained from 
the $i$th moment angular profile $\rho^{(i)}(\phi)$ as defined in Eq. (\ref{moments}).
 The magnitude 
$A_0(\kappa_1,\kappa_2,\phi)$ is directly related to the excluded area, 
$A_{\rm excl}(\kappa_1,\kappa_2,\phi)$, between two 
HR of aspect ratios $\kappa_1$ and $\kappa_2$ with a relative angle between their main axes  
equal to $\phi$. The relation between both magnitudes is 
$2A_0(\kappa_1,\kappa_2,\phi)=A_{\rm excl}(\kappa_1,\kappa_2,\phi)-\left(\kappa_1+\kappa_2\right)
\sigma^2$. In Fig. \ref{fig1} (a) we schematically show the excluded volume between two  
HR of $\kappa=2$ while in (b) we show the function $A_{\rm excl}(2,2,\phi)$. The  
secondary minimum at $\phi=\pi/2$ is responsible for the stability of the T phase at low aspect 
ratios which is a peculiar property almost unique to the rectangular shape. The most common    
functional form of the excluded area is exemplified for hard discorectangles (HDR) 
in (b) for comparison (note the presence of a maximum at $\pi/2$ instead of a local minimum).

\subsection{Coexistence calculations and numerical schemes}
\label{coexistence}

We now proceed to present the set of equations we used to calculate the phase coexistence between two phases with 
different orientational symmetries (I, N or T phases). The total free-energy density of a 
phase-separated system 
with a fraction $\gamma_{\alpha}$ of the total volume occupied by phase $\alpha$ 
($\alpha=\rm I,N,T$) can be 
expressed as $\Phi^{(\rm t)}[\rho]=\sum_{\alpha}\gamma_{\alpha} \Phi^{(\alpha)}[\rho]$ 
(obviously $\sum_{\alpha}\gamma_{\alpha}=1$). 

From now on the coexistence between (I,T)  and N phases occupying   
fractions of the total volume $\gamma_{I,T}=\gamma_0$ and $\gamma_{\rm N}=1-\gamma_0$ respectively will be 
denoted as (I,T)$_{\gamma_0}$-N$_{1-\gamma_0}$. For example I$_{0}$-N$_{1}$ coexistence implies that 
the I phase occupies a vanishing small volume (the so called shadow phase) while the N phase 
takes up the total volume (the so called cloud phase).

The lever rule guarantees 
the conservation of the total number of species of aspect ratio $\kappa$:   
\begin{eqnarray}
\rho_0f_0(\kappa)=\frac{1}{\pi}\sum_{\alpha}\gamma_{\alpha} \int_0^{\pi}d\phi
\rho^{(\alpha)}(\kappa,\phi),
\label{lever}
\end{eqnarray}
where $\rho_0=N/A$ (with $N$ the total number of particles) is the total number density of the system.
Using (\ref{lever}) as a constraint in the minimization of the total 
free-energy density $\Phi^{(\rm t)}[\rho]$ with respect to $\rho^{(\alpha)}
(\kappa,\phi)$ gives us the following set of equations:
\begin{eqnarray}
\rho^{(\alpha)}(\kappa,\phi)=\frac{\rho_0 f_0(\kappa)
\exp{\left[-\beta \mu_{\rm ex}^{(\alpha)}(\kappa,\phi)\right]}}
{\pi^{-1}\sum_{\tau} \gamma_{\tau}\int_0^{\pi} d\phi' 
\exp{\left[-\beta \mu_{\rm ex}^{(\tau)}(\kappa,\phi')\right]}}, 
\label{set2}
\end{eqnarray}
where $\alpha\in\{\rm I,T,N\}$ and the excess chemical potential of the coexisting $\alpha$-phase is 
defined as $\displaystyle{\beta\mu_{\rm ex}^{(\alpha)}(\kappa,\phi)=
\frac{\delta \Phi^{(\alpha)}_{\rm ex}[\rho]}{\delta\rho^{(\alpha)}
(\kappa,\phi)}}$. From now on we will use Greek letters $\alpha,\tau,\dots$ as superscripts 
to label coexisting phases while the Latin letters $i,j,\dots$ will be used to label the 
moments.

The numerical procedure used to solve Eqs. (\ref{set2}) is based on the reduction of this 
infinite set of equations (note that $\kappa$ and $\phi$ are continuous variables) in a finite set. 
To this purpose we firstly introduce a truncated Fourier expansion of the moment profiles: 
\begin{eqnarray}
\rho^{(i,\alpha)}(\phi)= \frac{1}{\pi}\sum_{j=0}^{N} \rho_j^{(i,\alpha)}\cos (2j \phi),
\label{Fourier}
\end{eqnarray}
with $\{\rho_j^{(i,\alpha)}\}$ the Fourier amplitudes. 
Secondly we substitute this expansion into 
$\beta\mu_{\rm ex}^{(\alpha)}(\kappa,\phi)$, multiply (\ref{set2}) by $\kappa^i\cos(2j\phi)$ 
and integrate over $\kappa$ and $\phi$ to obtain a closed set of equations: 
\begin{eqnarray}
&&\rho_j^{(i,\alpha)}=\rho_0D_{j0}\int_1^{\infty} d\kappa \kappa^i f_0(\kappa) 
\frac{T_j^{(\alpha)}(\kappa)}{\sum_{\tau} \gamma_{\tau} T_0^{(\tau)}(\kappa)}, \quad 
\label{set3}\\
&& T_j^{(\tau)}(\kappa)\equiv\pi^{-1} \int_0^{\pi} d\phi 
\exp{\left[-\beta\mu_{\rm ex}^{(\tau)}(\kappa,\phi)\right]}\cos(2j\phi),\nonumber\\
&&\label{las_T}
\end{eqnarray}
where $j=1,\dots,N,\ i=0,1$ and $D_{j0}=2/(1+\delta_{j0})$ with $\delta_{j0}$ the Kronecker delta. The success 
of the present strategy is based on the dependence of the excess part of the 
chemical potential on the angular density profile $\rho(\kappa,\phi)$ only through 
its moments $\{\rho_j^{(i,\alpha)}\}$: 

\begin{eqnarray}
&&\beta\mu_{\rm ex}^{(\alpha)}(\kappa,\phi)=-\ln\left[1-\rho_0^{(1,\alpha)}\sigma^2\right]+
\frac{S_1^{(\alpha)}(\kappa,\phi)}{1-\rho_0^{(1,\alpha)}\sigma^2}\nonumber\\
&&+\beta p^{(\alpha)} \kappa\sigma^2,\label{chepo}\\
&&S_1^{(\alpha)}(\kappa,\phi)\equiv \frac{2\sigma^2}{\pi}\left\{
\left(\rho_0^{(1,\alpha)}+\rho_0^{(0,\alpha)}\right)\left(\kappa+1\right)\right.\nonumber\\&&\left.-\sum_{j\geq 1} 
\frac{\left(\rho_j^{(1,\alpha)}+(-1)^j\rho_j^{(0,\alpha)}\right)\left(\kappa+(-1)^j\right)}{4j^2-1}\cos(2j\phi)\right\}.
\nonumber\\&&\\
&&\beta p^{(\alpha)}=
\frac{\rho_0^{(0,\alpha)}}{1-\rho_0^{(1,\alpha)}\sigma^2}+\frac{S_0^{(\alpha)}[\rho]}
{\left[1-\rho_0^{(1,\alpha)}\sigma^2\right]^2}, \label{pressure}\\
&&S_0^{(\alpha)}[\rho]=\frac{\sigma^2}{\pi}\left[\left(\rho_0^{(1,\alpha)}+\rho_0^{(0,\alpha)}\right)^2\right.\nonumber\\ 
&&\left.-\frac{1}{2}\sum_{j\geq 1}\frac{\left(\rho_j^{(1,\alpha)}+(-1)^j\rho_j^{(0,\alpha)}\right)^2}
{4j^2-1}\right]
\end{eqnarray}
where $\beta p^{(\alpha)}$, defined in Eq. (\ref{pressure}), is the pressure of phase $\alpha$.
Thus we finally obtain a closed set of equations (\ref{set3}) to be solved for the  
Fourier amplitudes $\{\rho_j^{(i,\alpha)}\}$. This set in turn guarantees the equality between 
chemical potentials of each species in both coexisting phases.   
An equation corresponding to the equality between pressures at each phase 
should be added to get the mechanical equilibrium, and consequently the pressure in 
Eq. (\ref{chepo}) can be dropped during the numerical implementation of (\ref{set3}). 
A fixed point iteration method was used to solve (\ref{set3}). To perform the numerical 
integrals over 
$\phi$ and $\kappa$ in Eqs. (\ref{set3}) and (\ref{las_T}) we used, after some convenient changes of variables, 
a Gauss-Legendre and Gauss-Laguerre quadratures respectively. Once the numerical 
convergence of Eqs. (\ref{set3}) was reached for a fixed $\rho_0$, we used the equality 
\begin{eqnarray}
p^{(\alpha)}\left(\{\rho_j^{(i,\alpha)}\},\rho_0\right)=p^{(\beta)}\left(\{\rho_j^{(i,\beta)}\},
\rho_0\right),
\label{mech}
\end{eqnarray}
numerically solved using the Brent's zero-finding method, to find the value of $\rho_0$ 
at equilibrium.   

\subsection{Bifurcation analysis}
\label{bifurca}

To find the onset of N and T ordering from the I phase we performed a bifurcation
analysis of Eqs. (\ref{set3}) particularized for $\gamma_{\rm I}=1$  
with respect to the small Fourier amplitudes $\{\rho^{(i,\alpha)}_j\}$ 
($j=1$ for $\alpha=$N while $j=2$ for $\alpha=$T). The resulting packing 
fractions, $\eta_0\equiv \rho_0 \kappa_0 \sigma^2$, at which the I phase is destabilized 
with respect to N and T phases (the so called spinodal instability) result in 
the following analytical expressions:
\begin{eqnarray}
&&\eta_0^{(\rm I-N)}=\left\{1+\frac{2}{3\pi\kappa_0}(\kappa_0-1)^2\left(1+\Delta_0^2\right)
\right\}^{-1},\\
&&\eta_0^{(\rm I-T)}=\left\{1+\frac{2}{15\pi\kappa_0}\left[(\kappa_0+1)^2+(\kappa_0-1)^2\Delta_0^2
\right]\right\}^{-1}.\nonumber\\&&
\end{eqnarray}

For small (large) mean aspect ratios $\kappa_0$ it is expected that the I phase 
destabilizes firstly with respect to the T (N) phase. Consequently there should exist a crossing 
point $\kappa_0^*$ between both, I-T and I-N, spinodals. This point can be calculated 
from the equality $\eta_0^{(\rm I-N)}=\eta_0^{(\rm I-T)}$ which results in 
\begin{eqnarray}
\kappa_0^*=\frac{3+2\Delta_0^2+\sqrt{5+4\Delta_0^2}}{2(1+\Delta_0^2)}.
\end{eqnarray}
From this equation we obtain, for the one-component fluid ($\Delta_0=0$), the value  
$\kappa_0^*=1+\varphi$ with $\varphi=(1+\sqrt{5})/2\approx 1.618$ the golden ratio. 
For the highest polydispersity, $\Delta_0=1$ 
(reached for the Schulz distribution), we obtain $\kappa_0^*=2$. Thus we can extract 
as a preliminary conclusion that  the stability of T phase slightly decreases with polydispersity. 

For small aspect ratios, $\kappa_0\sim 1$, it is also expected an stable T phase
up to the density at which a T-N transition takes place. To calculate the T-N spinodal we have 
performed a bifurcation analysis of Eqs. (\ref{set3}), particularized for $\gamma_{\rm T}=1$, 
with respect to the small Fourier components $\{\rho_1^{(i,\rm N)}\}$. The obtained result for 
the packing fraction at the N-T bifurcation can be calculated from
\begin{eqnarray}
&&\eta_0^{(\rm N-T)}=\left\{1+\frac{2}{3\pi\kappa_0}\left[(\kappa_0-1)^2\left(1+\Delta_0^2\right)
\right.\right.\nonumber\\&&\left.\left.+\chi_2\left(\eta_0^{(N-T)}\right)\right]\right\}^{-1},
\label{itera}
\end{eqnarray}
where we have defined 
\begin{eqnarray}
\chi_2(\eta_0)=\int d\kappa f_0(\kappa)(\kappa-1)^2 \frac{T_2^{(\rm T)}(\kappa)}
{T_0^{(\rm T)}(\kappa)},
\end{eqnarray}
with the functions $T_j^{(T)}(\kappa)$, already defined in (\ref{las_T}), calculated 
once the Fourier amplitudes $\{\rho_{2j}^{(i,\rm T)}\}$ at the equilibrium T phase were 
found for a given $\eta_0$. We have solved iteratively the nonlinear integral equation (\ref{itera})  
with respect to $\eta_0^{(\rm N-T)}$ to find the N-T spinodal instability. 

\begin{figure}
\epsfig{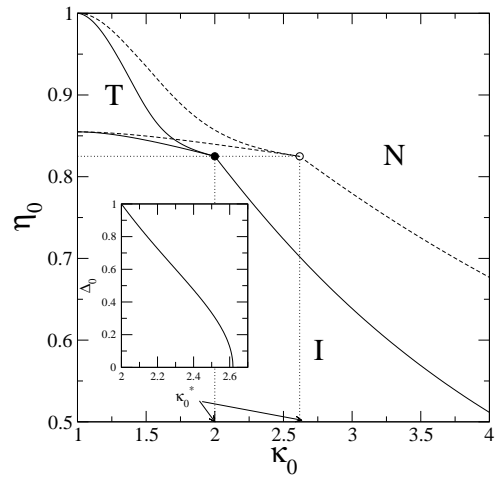}
\caption{I-N, I-T and T-N spinodal curves: packing fraction  vs. mean aspect ratio 
$\kappa_0$ for $\Delta_0=0$ (dashed) and $\Delta_0=1$ (solid). 
Inset: Polydisperse coefficient $\Delta_0$ as a function of 
the critical value $\kappa_0^*$ for which $\eta_0^{(\rm I-N)}=\eta_0^{(\rm I-T)}$ 
(the filled and open circles in the main figure). The stability regions 
of I, N and T phases are correspondingly labeled.}
\label{fig2}
\end{figure}

We plot the three found spinodals $\eta_0^{(\rm I-N)}(\kappa_0)$, $\eta_0^{(\rm I-T)}(\kappa_0)$ and 
$\eta_0^{(\rm N-T)}(\kappa_0)$ in Fig. \ref{fig2} for $\Delta_0=0$ and 1. 
We can see that the I-N spinodal is severely affected by polydispersity 
(is significantly below its one-component counterpart). 
Another conclusion we can extract from the figure is that 
the stability of T phase only slightly decreases with polydispersity  
as we have already point out before by comparing the values of the crossing points 
$\kappa_0^*$.  
This destabilization can be explained by the presence of long rods (those which destroy the 
T symmetry) with aspect ratios much larger than $\kappa_0$. The long rods      
have a profound effect on the orientational ordering properties of the fluid, usually favoring the N ordering. 
However the decrease in the T phase stability is not so strong as expected because 
the mean-square deviation, $\Delta$,  of the function $f_0(\kappa)$ 
decreases with the mean aspect ratio $\kappa_0$ 
[see Eq. (\ref{deviation}) ]: when $\kappa_0\to 1$ we obtain $\Delta \to 0$ for any $\Delta_0$.   

\subsection{Measuring the orientational ordering and fractionation effects}
\label{measures}
To present the results in Sec. \ref{Results} it is convenient to define, 
apart from the already introduced packing fraction $\eta_0$, 
the following dimension-less variables: 
\begin{eqnarray}
\eta_j^{(0,\alpha)}= \rho_j^{(0,\alpha)}\kappa_0\sigma^2,\quad  
\eta_j^{(1,\alpha)}=\rho_j^{(1,\alpha)}\sigma^2.
\end{eqnarray}

To measure the orientational ordering of rectangles of aspect ratio $\kappa$ 
at the coexisting $\alpha$-phase we use the angular probability distribution function:
\begin{eqnarray}
h^{(\alpha)}(\kappa,\phi)\equiv \frac{\rho^{(\alpha)}(\kappa,\phi)}{\int_0^{\pi}d\phi' \rho^{(\alpha)}(\kappa,\phi')}.
\label{meassure}
\end{eqnarray}
The integration of $h^{(\alpha)}(\kappa,\phi)$ over $\kappa$ gives us the averaged
angular distribution function which, 
resorting to the Fourier expansion (\ref{Fourier}), can be easily calculated using the expression:
\begin{eqnarray}
h^{(\alpha)}(\phi)=\left(\pi \rho_0^{(0,\alpha)}\right)^{-1}\sum_j \rho_j^{(0,\alpha)}\cos(2j\phi).
\label{angular0}
\end{eqnarray}
The N ($j=1$) and and T ($j=2$) order parameters are defined through 
\begin{eqnarray}
Q_j^{(\alpha)}=\int_0^{\pi}d\phi h^{(\alpha)}(\phi)\cos(2j\phi).
\label{param}
\end{eqnarray}
Inserting (\ref{angular0}) into (\ref{param}) we obtain
\begin{eqnarray}
Q_j^{(\alpha)}=\frac{\rho_j^{(0,\alpha)}}{2\rho_0^{(0,\alpha)}}.
\label{ordering}
\end{eqnarray}

To measure the fractionation between the cloud and shadow coexisting phases 
we should bear in mind that the cloud 
phase always has the parent distribution function $f_0(\kappa)$. Thus we will concentrate only on the shadow-phase 
distribution functions. For the I$_1$-N$_0$ and I$_0$-N$_1$ coexistences the distributions 
corresponding to the N and I shadow phases are 
\begin{eqnarray}
&&f^{(\rm N)}(\kappa)=f_0(\kappa)e^{\beta\mu_{\rm ex}^{(\rm I)}(\kappa)} T_0^{(\rm N)}(\kappa), \\
&&f^{(\rm I)}(\kappa)=f_0(\kappa)\frac{e^{-\beta\mu_{\rm ex}^{(\rm I)}(\kappa)}}{T_0^{(\rm N)}(\kappa)},
\label{plotear1}
\end{eqnarray}
while those corresponding to N and T shadow phases of the T$_1$-N$_0$ and T$_0$-N$_1$ coexistences are
\begin{eqnarray}
&&f^{(\rm N)}(\kappa)=f_0(\kappa) \frac{T_0^{(\rm N)}(\kappa)}{T_0^{(\rm T)}(\kappa)},\\ 
&&f^{(\rm T)}(\kappa)=f_0(\kappa)\frac{T_0^{(\rm T)}(\kappa)}{T_0^{(\rm N)}(\kappa)}.
\label{plotear2}
\end{eqnarray}
All these distributions will be normalized as $\tilde{f}^{(\alpha)}(\kappa)\equiv f^{(\alpha)}
(\kappa)/\int d\kappa f^{(\alpha)}(\kappa)$ for plotting. We will use the averaged  
over $\tilde{f}^{(\alpha)}(\kappa)$ aspect ratio,
\begin{eqnarray}
\langle \kappa\rangle_{\tilde{f}^{(\alpha)}}\equiv \int d\kappa \kappa \tilde{f}^{(\alpha)}(\kappa)=
\frac{\eta_0^{(1,\alpha)}}{\eta_0^{(0,\alpha)}},
\end{eqnarray} 
to quantify the fractionation. Note that it is equal to $\kappa_0$ when averaged with respect 
to $f_0(\kappa)$.

\section{Results}
\label{Results}
In this section we show the results as obtained from the numerical implementation of SPT-DF for length-polydisperse 
HR. We divide this section in different subsections each one devoted to discuss the thermodynamic 
and orientational properties of polydisperse HR. In Sec. \ref{phd} we firstly describe the resulting phase  diagrams 
using two different parent distribution functions (those with exponential and Gaussian-like decays). 
Further we proceed, in Sec. 
\ref{orient}, to present a quantitative analysis of the orientational ordering of particles at coexistence. To this 
purpose we use the angular distribution functions and order parameters. The fractionation of species with 
different aspect ratios between coexisting phases as a function of the mean aspect ratio and 
polydispersity is studied in Sec. \ref{frac1} and \ref{frac2} respectively. 
Finally, in Sec. \ref{tricri} 
we study the effect of polydispersity on the location of the I-N tricritical point.  

\subsection{Phase diagrams for fixed polydispersity $\Delta_0$}
\label{phd}

\begin{figure}
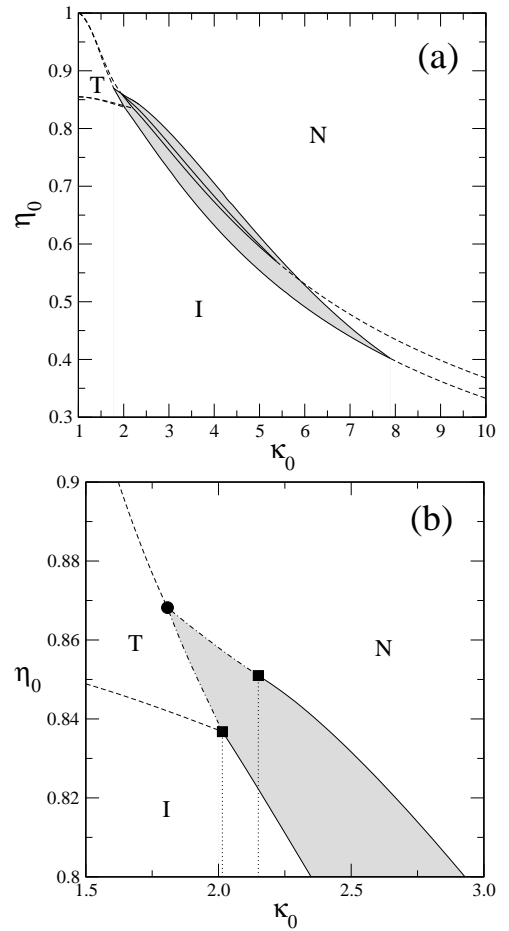

\epsfig{file=fig3a.eps,width=2.5in}
\epsfig{file=fig3b.eps,width=2.5in}
\caption{ 
(a): Phase diagrams of the one-component ($\Delta_0=0$) and polydisperse ($q=1$, and $\Delta_0=0.408$) HR fluid in the 
total packing fraction $\eta_0$ vs. mean aspect ratio $\kappa_0$ plane. The one-component fluid 
exhibits a weaker first order I-N transition finishing at a tricritical point located well bellow that of the 
polydisperse fluid. With solid and dashed lines are shown the coexistence binodals and spinodals respectively. 
Regions of stability of I, N and T phases are labeled in the figure. (b): A zoom of the phase diagram of 
polydisperse HR shown in (a) around the left tricritical point (solid circle). With dot-dashed lines are shown 
the T$_1$-N$_0$ and T$_0$-N$_1$ coexistence binodals both ending at the solid squares. Between the vertical 
dotted lines a possible triple I-N-T coexistence could exist.}
\label{fig3}
\end{figure} 

We numerically solved the set of equations (\ref{set3}) together with the mechanical equilibrium condition 
(\ref{mech}) selecting a parent distribution function (\ref{distri0}) with $q=1$ and a polydisperse 
coefficient $\Delta_0=0.408$, corresponding to the value $\nu=5$ in Eq. (\ref{distri0}). 
For small mean aspect ratios, $\kappa_0\sim 1$, we calculated the T-N coexistence from Eqs. (\ref{set3}) and (\ref{mech})
fixing the odd Fourier coefficients $\rho_{2j+1}^{(i,\rm T)}$ to zero. 
This  simplification is justified by the symmetry 
of the T phase: $\rho^{(\rm T)}(\kappa,\phi)=\rho^{(\rm T)}(\kappa,\phi+\pi/2)$. For medium and large aspect ratios 
we solved the same set of equations taking into account all the Fourier coefficients $\rho_{j}^{(i,\rm N)}$ 
to look for the I-N coexistence. 
We have found that the I-T transition is always of second order.
The total packing fractions $\eta_0$ of the I$_1$ and N$_1$ (coexisting with their shadow counterparts) phases are 
shown in Fig. \ref{fig3}. Together with these coexisting binodals we also plot the I-N, I-T and N-T 
spinodals as calculated in Sec. \ref{bifurca}. For comparison we also plot in (a) the phase diagram corresponding 
to the one-component HR fluid ($\Delta_0=0$) calculated in Ref. \cite{M-R6}. 
The most prominent effects of polydispersity consist 
of: (i) the dramatic broadening of the I-N coexisting region, (ii) the displacement of the I-N tricritical point to 
higher values of $\kappa_0$, and (iii) the lowering of the second-order I-N transition density (the spinodal curve). 
The widening of the two-phase coexistence region is a well known fact in many studies 
conducted on three-dimensional polydisperse hard-rods \cite{Speranza1,Speranza2,Speranza3,Wensink}. 
As it is well known the I-N transition of hard rods in 3D is of first 
order for any aspect ratio. In 2D the mean-field DF predicts a second order 
I-N transition for most particle shapes, HR being an important exception: The I-N and T-N transitions 
are of first order for aspect ratios between two (N-T and I-N) tricritical points. The broadening of the I-N 
coexisting region found here is related, in analogy to 3D, to a demixing-like mechanism: the N and I phases 
are preferentially populated by long and small rods respectively (the so called fractionation effect). 
We can see from Fig. \ref{fig3} that the T phase slightly destabilizes with polydispersity as we have already discussed 
in Sec. \ref{bifurca}. 

\begin{figure}
\epsfig{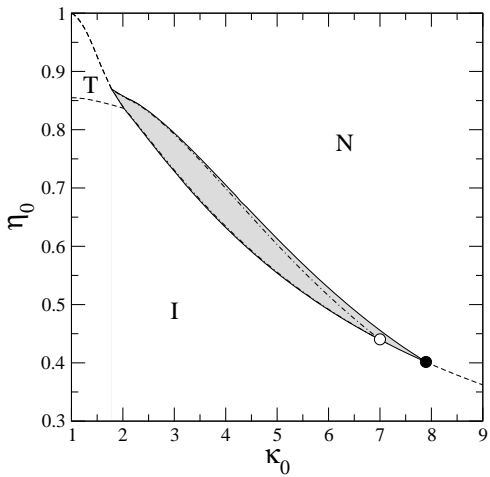}
\caption{Phase diagrams for $\Delta_0=0.408$, $q=1$ (solid), and $q=2$ 
(dot-dashed). With solid and empty circles are shown the I-N tricritical points corresponding to $q=1$ and 2 
respectively.} 
\label{fig4}
\end{figure}

In Fig. \ref{fig3} (b) we show a zoom of the phase diagram around the left T-N tricritical 
point (solid circle). Note that the T$_1$-N$_0$ binodal (the lower dot-dashed curve) ends up 
at an aspect ratio value $\kappa_0$ (lower square) slightly below that corresponding to the  
end of the T$_0$-N$_1$ binodal (upper square). This 
result suggests that the system could exhibit a triple I-T-N coexistence for aspect ratios  
located between the two vertical dotted lines. 

\begin{figure*}
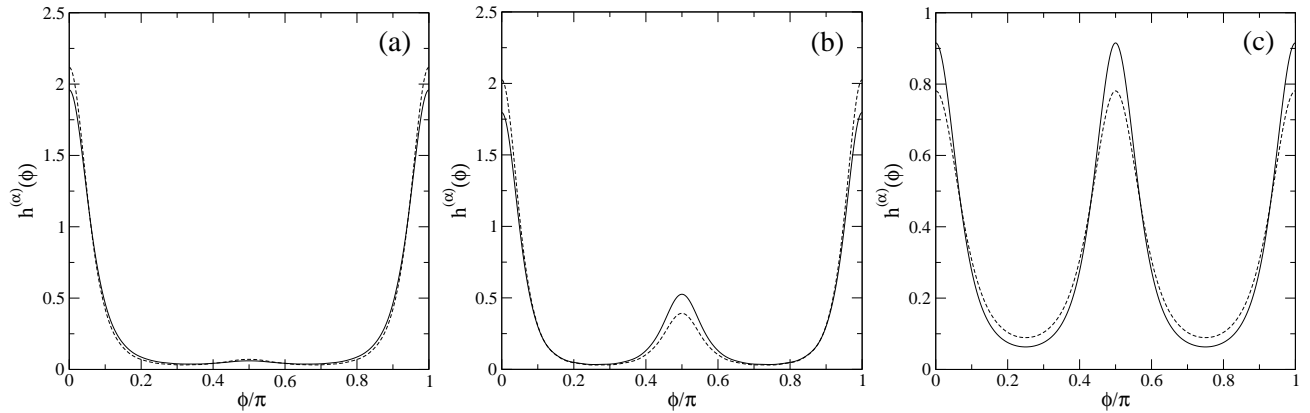

\epsfig{file=fig5a.eps,width=2.2in}
\epsfig{file=fig5b.eps,width=2.2in}
\epsfig{file=fig5c.eps,width=2.2in}
\caption{Angular distribution functions $h^{(\alpha)}(\phi)$ 
for $\alpha$=(N,T)$_1$ (solid) and $\alpha=$(N,T)$_0$ (dashed) phases corresponding to the I-N coexistence for 
$\kappa_0=3.2$ (a) and to the T-N coexistence for $\kappa_0=1.9$ (b) and (c). In (b) and (c) we show 
$h^{(\rm N)}(\phi)$ and $h^{(\rm T)}(\phi)$ respectively. The values of $q$ and $\Delta_0$ were fixed 
to 1 and 0.4068.}   
\label{fig6}
\end{figure*}

The phase diagram corresponding to the selected parent distribution function with $q=2$ (that with a Gaussian decay)
is shown in Fig. \ref{fig4} for the same $\Delta_0=0.408$. 
For comparison we show in the same figure the case $q=1$. We observe that the cloud binodals corresponding to the 
I$_1$-N$_0$ coexistences for $q=1$ and 2 coincide.
On the other hand the cloud binodal of the 
of the I$_0$-N$_1$ coexistence for $q=2$ is located, in particular for large aspect ratios, 
below that of $q=1$, reducing in this way the coexisting gap and consequently moving the I-N tricritical 
point from $\kappa_0^{(\rm t)}\approx 7.9$  to $\kappa_0^{(\rm t)}\approx 7.05$. We can conclude from 
this result that the position of the tricritical point depends not only 
of the first two moments of the parent distribution function (both are the same for $q=1$ and 2), but also on  
higher moments which are certainly different for the Schulz and Gaussian-like distributions. Through a 
bifurcation analysis of the free-energy with respect to the Fourier components of an incipient N phase 
performed around the tricritical point we can obtain an analytic equation relating the position of this point 
with these moments. This analysis was done for binary mixtures 
of HR in Ref. \cite{M-R8} and we plan to extend it to the polydisperse case in a future. The reduction of 
the coexisting gap with $q$ reflects the lesser relevance of long rods on the phase behavior for  
distributions with strong decay.   

\subsection{Orientational ordering}
\label{orient}

In this section we discuss the orientational ordering of rectangles at equilibrium. To this purpose 
we firstly show the coexisting angular distribution functions $h^{(\alpha)}(\phi)$ as calculated 
from Eq. (\ref{angular0}) corresponding to an stable I-N phase separation for 
$\kappa_0=3.2$, $q=1$ and $\Delta_0=0.408$. The cloud and shadow N distributions are plotted in 
Fig. \ref{fig6} (a) with solid and dashed lines respectively. The sharp peaks at 0 and $\pi$ show 
the strong N ordering present in the fluid, while a rather small oscillation around $\pi/2$ 
indicates the closeness of the system to the region of T phase stability ($1\leq \kappa_0\lesssim 2$). 
Also, the cloud and shadow N distributions are very similar, with the
latter showing an slightly higher ordering of particles. We show these distributions 
for the same values of $q$ and $\Delta_0$ but this time selecting $\kappa_0=1.9$ where the T-N 
coexistence is stable. The results are shown in Fig. \ref{fig6} (b) and (c) for the 
N and T phases respectively. Now the N phase develops a well defined central peak at $\pi/2$ 
reflecting a high proportion of small rectangles with T-like ordering. Again 
the shadow N phase exhibits slightly higher degree of N ordering and consequently  
a lesser proportion of small rectangles with T-like configurations. The coexisting T 
distributions [see panel (c)] has a periodicity of $\pi/2$ (the central peak is of the same 
height as those located at 0 and $\pi$) reflecting the T symmetry: the system is invariant under 
rotations of $\pi/2$. The cloud T phase has a higher degree of ordering than its shadow counterpart. 

\begin{figure}
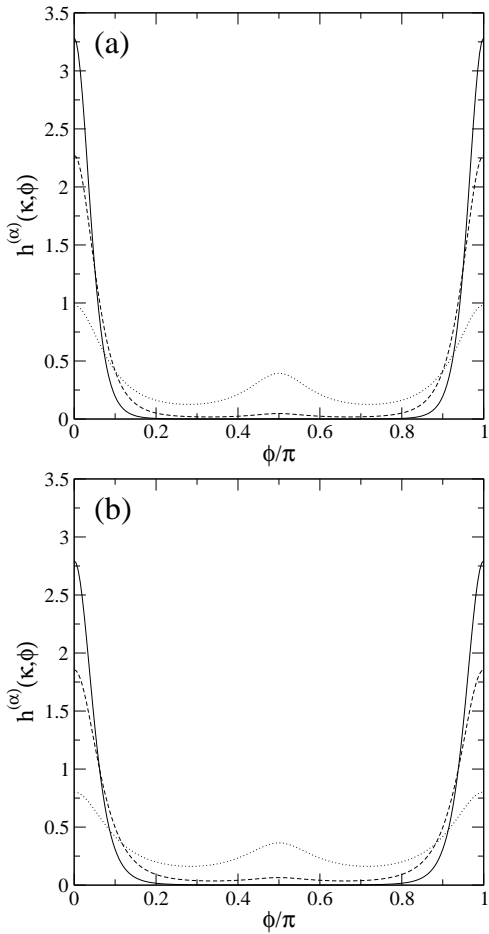

\epsfig{file=fig6a.eps,width=2.5in}
\epsfig{file=fig6b.eps,width=2.5in}
\caption{Orientational distribution functions of rectangles of 
aspect ratios  $\kappa= 1.5$ (dotted), 3 (dashed) and 5 (solid)
corresponding to the I$_0$-N$_1$ (a) and $I_1$-N$_0$ (b) coexistences of a polydisperse mixture 
with $q=1$, $\kappa_0=3$ and $\Delta_0=0.408$.}
\label{fig8}
\end{figure}

A better understanding of the degree of particle ordering at I-N coexistence can be reached through 
the calculation of the angular distribution function of species of a certain aspect ratio $\kappa$, 
$h^{(\alpha)}(\kappa,\phi)$, as defined by Eq. (\ref{meassure}). It is expected that 
the N (T) ordering 
increases (decreases) with $\kappa$, an assertion well supported by the  
N distributions shown in Fig. \ref{fig8} and calculated for the parameters 
$q=1$, $\Delta_0=0.408$, $\kappa_0=3$, and for three different values of $\kappa$: 
1.5 (dotted), 3 (dashed) and 5 (solid). The small species, being the less ordered, present 
a high proportion of T-like configurations while the large ones exhibit 
a high degree of uniaxial N ordering. Again the  cloud distributions [panel (a)] 
reflect a higher ordering than their shadow counterparts [panel (b)], a fact already pointed out 
before when we discussed the behavior of $h^{(\alpha)}(\phi)$ for $\kappa_0=3.2$.

\begin{figure}
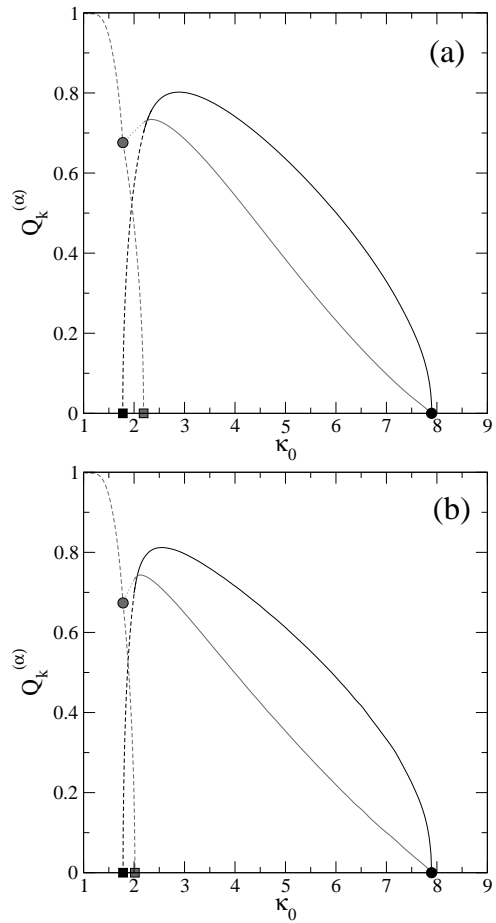

\epsfig{file=fig7a.eps,width=2.5in}
\epsfig{file=fig7b.eps,width=2.5in}
\caption{Order parameters $Q_j^{(\alpha)}$ ($j=1,2$, and $\alpha=\rm N,T$) as a function of $\kappa_0$ 
for $\gamma_{\rm N}=0$ (a) and $\gamma_{\rm N}=1$ (b). Different lines represent 
the order parameters along the I-N (solid lines) or T-N (dashed lines) coexistences. 
The order parameters $Q_1^{(\alpha)}$ and $Q_2^{(\alpha)}$ are shown in black and grey 
respectively. With grey and black solid circles are shown the left T-N and right I-N tricritical 
points while the squares represent the values of $\kappa_0$ where $Q_2^{(\alpha)}=0$.} 
\label{fig5}
\end{figure}

The global ordering along the coexisting binodal and spinodal curves of 
the phase diagram, shown 
in Fig. \ref{fig3}, is calculated through the order parameters $Q_{1,2}^{(\alpha)}$ 
[see Eq. (\ref{ordering})].  
These are plotted in Fig. \ref{fig5} as functions of $\kappa_0$. For large values of $\kappa_0$, not 
close enough to the I-N tricritical point, $\kappa_0^{t_2}$,   
the N order parameter $Q_1^{(\rm N)}$ 
along the I-N coexistence is relatively high and always above the T order parameter 
$Q_2^{(\rm N)}$. 
Obviously at the I-N tricritical point they both become zero. 
As $\kappa_0$ decreases from $\kappa_0^{t_2}$ the T and N order parameters 
increase, both reaching their maxima at slightly different values close to $\kappa_0=3$. Just 
below $\kappa_0=3$ a cross-over  
between $Q_1^{(\rm N)}$ and $Q_2^{(\rm N)}$ takes place.
A slight decrease of $\kappa_0$ from this cross-over  
gives $Q_2^{(\rm N)}>Q_1^{(\rm N)}$, a clear sign of the presence of a secondary 
peak at $\pi/2$ in the angular distribution function. Then the I-N transition is substituted 
by the T-N one and $Q_1^{(\rm N)}$ decreases to zero at the T-N tricritical point $\kappa_0^{t_1}$. 
The order parameter $Q_2^{(\rm N)}$  also decreases in a lesser extent up to the value 
shown with a grey circle (the T-N tricritical point). 
From this point 
$Q_2^{(\rm T)}$ increases along the T-N spinodal  
up to its maximum allowed value 1 as $\kappa_0\to 1$ while it decreases along the 
T-N coexistence binodal as $\kappa_0$ is increased up to $\kappa_0^c$, the value 
corresponding to the end-critical point, where the I-T spinodal and the cloud binodal of the T$_1$-N$_0$ coexistence  
meet.  
The difference, $\Delta\kappa\equiv \kappa_0^c-\kappa_0^{t_1}$, between the aspect ratios 
corresponding 
to those of the end-critical and tricritical points [see the square symbols in (a) and (b)] 
depends on  
which of the phases is considered to be the cloud one. When $\gamma_{\rm N}=1$ [panel (a)], 
i.e. for the (I,T)$_0$-N$_1$ coexistence, this difference is clearly larger than that  
corresponding to $\gamma_{\rm N}=0$, the (I,T)$_1$-N$_0$ coexistence [panel (b)]. 
This behaviour is related to the already discussed fact on the possible existence of a triple I-N-T coexistence 
[see Fig. \ref{fig3} (b)] for values of mean aspect ratios located between these points.

\subsection{Fractionation as a function of $\kappa_0$}
\label{frac1}
\begin{figure*}
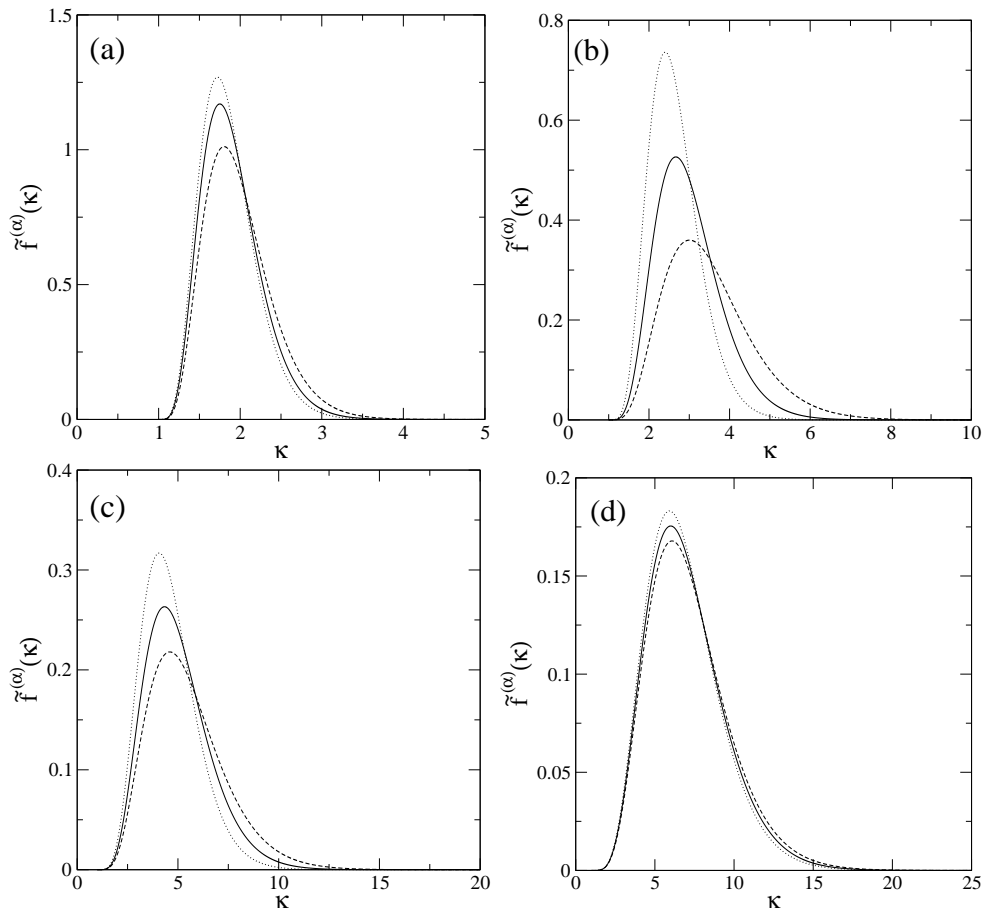

\epsfig{file=fig8a.eps,width=2.5in}
\epsfig{file=fig8b.eps,width=2.5in}
\epsfig{file=fig8c.eps,width=2.5in}
\epsfig{file=fig8d.eps,width=2.5in}
\caption{Distribution functions $\tilde{f}^{(\alpha)}(\kappa)$ of T-N (a) and I-N [(b)-(d)] 
coexisting phases for $\kappa_0=1.9$ (a), 3 (b), 5 (c), and 7 (d). The parent distribution function $f_0(\kappa)$ 
(coinciding with the distributions of the cloud  coexisting phases) was 
selected to be of Schultz-type ($q=1$) with a polydisperse coefficient $\Delta_0=0.408$ and it is plotted with  
solid lines. The distributions corresponding to N$_0$ (a-d) are plotted with dashed lines while 
those of T$_0$ (a) and I$_0$ (b-d) are plotted with dotted lines.}
\label{fig10}
\end{figure*}

This section is devoted to explore in detail the fractionation of different species between 
the coexisting phases. To this purpose we have fixed the values of the mean aspect 
ratio $\kappa_0$ to one of the values of the following set: $\{3,5,7\}$. Also we have selected   
the polydisperse coefficient of the Schulz-type ($q=1$) 
distribution function to be $\Delta_0=0.408$ as before. Further, we numerically 
solved the Eqs. (\ref{set3}) and (\ref{mech}) for chemical and mechanical equilibrium 
between cloud-shadow coexisting phases. 
As a result we found the normalized length-distributions, 
$\tilde{f}^{(\alpha)}(\kappa)$, obtained from (\ref{plotear1})-(\ref{plotear2}) for the 
shadow I, T or N phases. We should bear in mind that the cloud phases always follow 
the distribution $f_0(\kappa)$. All these distributions are plotted in Fig. \ref{fig10}. We can 
appreciate a clear fractionation of small/long species between cloud-shadow phases. The small 
ones preferentially populate the I or T phases while the N phase is rich in long species: see how the 
maxima of the I/T distributions are clearly located left 
of those of the N distributions. Also the latter has a slower decay, 
indicating a higher fraction of long species. As we can see from the figure the fractionation is much more
dramatic for $\kappa_0=3$ which it is clearly correlated with the value at which the N phase has the 
highest order parameter $Q_1^{(\rm N)}$ (see Fig. \ref{fig5}). As we move away from this value in   
the directions of both tricritical points $\kappa_0^{t_i}$ the system exhibits fractionation but in 
a lesser extent due to the weaker character of the (I,T)-N phase transition.   

\begin{figure}
\epsfig{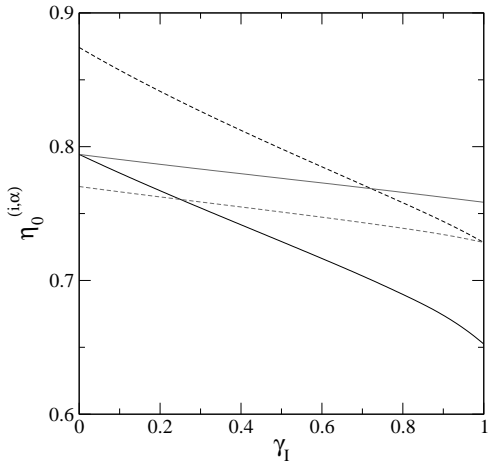}
\caption{Moments $\eta_0^{(0,\alpha)}$ (black) and $\eta_0^{(1,\alpha)}$ (grey) 
of the I (dashed) and N (solid) coexisting distribution functions as 
a function of the fraction of volume occupied by I phase, $\gamma_{\rm I}$, 
for $\kappa_0=3$, $q=1$ and $\Delta_0=0.408$.}
\label{fig7}
\end{figure}

Another important consequence of the fractionation is related to the values of the unit-less moments 
of the distribution functions, $\eta_0^{(i,\alpha)}$, at coexistence. 
In Fig. \ref{fig7} we plot these moments as a function of 
$\gamma_{\rm I}$ for $q=1$, $\kappa_0=3$ and $\Delta_0=0.408$. The inequalities 
$\eta_0^{(0,\rm I)}>\eta_0^{(0,\rm N)}$ and $\eta_0^{(1,\rm N)}> \eta_0^{(1,\rm I)}$ are always 
fulfilled which constitutes a direct consequence of how different are the shapes of 
$\tilde{f}^{(\rm I)}(\kappa)$ and $\tilde{f}^{(\rm N)}(\kappa)$ as discussed previously. Integrating 
the sharpest, strongly decayed function $\kappa^i \tilde{f}^{(\rm I)}(\kappa)$, 
over $\kappa$ for $i=0$ ($i=1$) gives us 
a value of $\eta_0^{(0,\rm I)}$ ($\eta_0^{(1,\rm I)}$) greater (less) than that 
obtained from the integration of the slower decaying function $\kappa^i \tilde{f}^{(\rm N)}(\kappa)$.

\begin{figure}
\epsfig{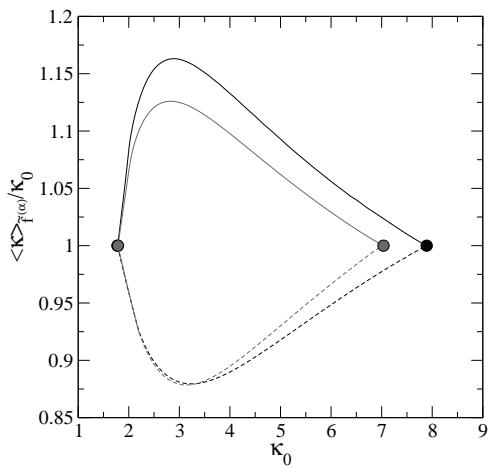}
\caption{Averaged over $\tilde{f}^{(\alpha)}(\kappa)$  aspect ratio, 
$\langle\kappa\rangle_{\tilde{f}^{(\alpha)}}$, 
in units of the mean aspect ratio $\kappa_0$ 
along the coexisting (I,T)$_0$ (dashed) and N$_0$ (solid) binodals for $q=1$ (black) and $q=2$ (grey).  
The polydispersity is fixed to $\Delta_0=0.408$.} 
\label{fig11}
\end{figure}

To finish this section we present in Fig. \ref{fig11} how  
$\langle \kappa\rangle_{\tilde{f}^{(\alpha)}(\kappa)}$ (the averaged aspect ratio with respect 
to the coexisting shadow distribution $\tilde{f}^{(\alpha)}(\kappa)$) evolves as a function of $\kappa_0$ 
for $\Delta_0=0.408$. As expected, this average, which directly measures
the fractionation, becomes equal to $\kappa_0$ at both tricritical points (solid circles) while 
it reaches its maximum (minimum) value at $\kappa_0\sim 3$ when averaged with respect to  
N$_0$ (I$_0$) distributions. This qualitative behavior is similar for both, the Schulz ($q=1$) 
and Gaussian-tailed ($q=2$) distributions. However the latter provokes less fractionation.

\subsection{Fractionation as a function of $\Delta_0$}
\label{frac2}
\begin{figure}
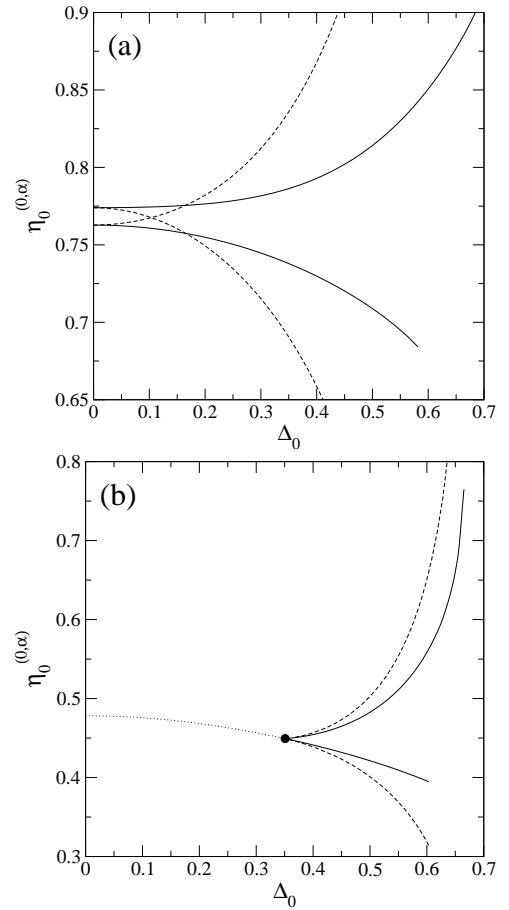

\epsfig{file=fig11a.eps,width=2.5in}
\epsfig{file=fig11b.eps,width=2.5in}
\caption{The zeroth moments $\eta_0^{(0,\alpha)}$ ($\alpha=\rm{I,N}$) as a function of $\Delta_0$ for $q=1$ and 
$\kappa_0=3$ (a) and $\kappa_0=7$ (b). With solid 
lines are shown the values of $\eta_0^{(0,\alpha)}$ for the I$_1$ and N$_1$ coexisting (with N$_0$ and I$_0$) 
phases respectively. The (I,N)$_0$ coexisting curves are shown with dashed lines. The black circle in (b) shows 
the position of the tricritical point while the dotted line represents the I-N spinodal curve.} 
\label{fig9}
\end{figure}

\begin{figure}
\epsfig{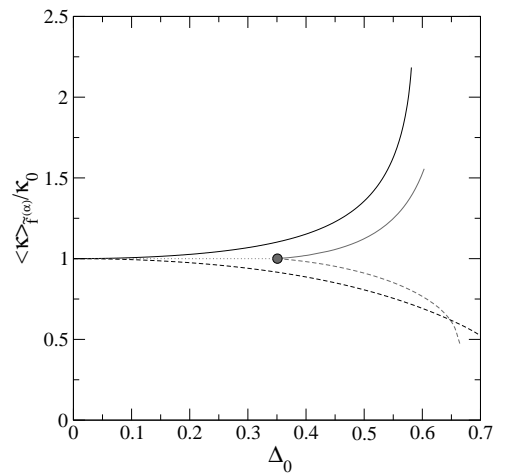}
\caption{Averaged over $\tilde{f}^{(\alpha)}(\kappa)$  [$\alpha=$I (dashed) and $\alpha=$N (solid)] 
aspect ratio, $\langle\kappa\rangle_{\tilde{f}^{(\alpha)}}$, 
in units of $\kappa_0$ as a function of $\Delta_0$ for $\kappa_0=3$ (black) and 7 (grey) and $q=1$.}
\label{fig12}
\end{figure}

To finish the discussion on fractionation we present in Fig. \ref{fig9} the results 
regarding the behaviour of the 
dimension-less zeroth moments, $\eta_0^{(0,\alpha)}$, 
of coexisting cloud and shadow phases as a function of polydispersity for fixed 
$\kappa_0=3$ [(a)] and 7 [(b)] and selecting $f_0(\kappa)$ to be of Schulz type ($q=1$). 
We observe a dramatic widening of the coexistence region, 
i.e. the gap between the moments corresponding to I$_1$ and N$_1$ phases (solid lines) greatly 
increases with $\Delta_0$ (for $\kappa_0=7$ the same occurs but 
beyond the tricritical point). Also, for 
$\kappa_0=3$ and zero polydispersity the moments of the shadow I and N phases (dashed lines) 
obviously coincide with those of the cloud phases and they compare as usually: 
$\eta_0^{(0,\rm I)}<\eta_0^{(0,\rm N)}$.  However as $\Delta_0$ increases they exhibits a cross-over 
at $\Delta_0\approx 0.1$ and the former relation inverts: $\eta_0^{(0,\rm I)}>\eta_0^{(0,\rm N)}$. 
This in turns means that the I shadow distribution function $f^{(\rm I)}(\kappa)$ (that which it is 
not normalized) exhibits, as a result of strong fractionation, a sharper peak located at relatively 
small values of $\kappa$. On the other hand the N shadow distribution, 
$f^{(\rm N)}(\kappa)$, shows a more smeared peak located at larger values of $\kappa$ 
(see Fig. \ref{fig10}). These differences in the shapes of distributions result in 
$\int d\kappa f^{(\rm I)}(\kappa)> \int d\kappa f^{(\rm N)}(\kappa)$. 
Interestingly this cross-over does not exist for 
$\kappa_0=7$:  At the tricritical point the moments are equal while beyond 
it the fractionation mechanism always gives 
$\eta_0^{(0,\rm I)}>\eta_0^{(0,\rm N)}$.
A more clear measure of fractionation, given by 
$\langle \kappa\rangle_{\tilde{f}^{\alpha}}$, 
is plotted as a function of $\Delta_0$ along the same shadow curves in Fig. \ref{fig12}. For 
$\kappa_0=3$ (7) the aspect ratio, averaged with respect to the shadow N (I) distribution,  
can reach values as large (low) as $2\kappa_0$ ($\kappa_0/2$) for high enough $\Delta_0$, 
a clear sign of the presence of strong fractionation.

\subsection{Tricritical points}
\label{tricri}
\begin{figure}
\epsfig{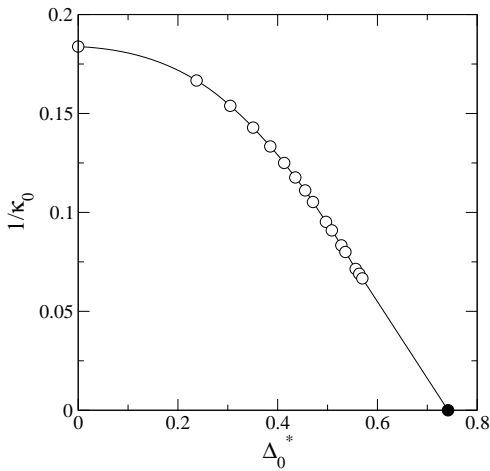}
\caption{$\kappa_0^{-1}$ as a function of $\Delta_0^*$ (the I-N tricritical point value) for $q=1$. 
The open circles 
correspond to the calculated values while the solid curve is an Akima cubic-spline fitting. The black circle 
indicates the extrapolated value of $\Delta_0^*$ when $\kappa_0\to\infty$.}  
\label{fig13}
\end{figure}

The last study we have carried out concerns the location of the I-N tricritical point as 
a function of polydispersity. To this purpose we fixed the value of $\kappa_0$ and 
calculated the I$_0$-N$_1$ coexistence for high enough values of $\Delta_0$ [selected 
in such a way to guarantee a first order I-N transition, and thus to find a nontrivial 
numerical solution of Eqs. (\ref{set3})]. Then we gradually 
decrease the value of $\Delta_0$  and used as new initial guesses  
those found in the previously converged iterations. This 
process is repeated up to that value of $\Delta_0$ for which the order parameter $Q_1^{(\rm N)}$ 
is negligibly small. Finally an accurate extrapolation 
(using a cubic-spline fitting) of $Q_1^{(\rm N)}$ to zero gives us the value $\Delta_0^*$ at 
which the first order transition becomes of second order, i.e. the position of the tricritical 
point. This process was carried out for a set of aspect ratios and the results are shown 
in Fig. \ref{fig13}. The open circles show our selected values of $\kappa_0$ and the curve 
is an Akima spline fitting to guide the eyes. This curve has certain 
credibility only as an interpolation fitting. However we have decided to 
use the same fitting to conjecture the value of $\Delta_0^*$ in the limit $\kappa_0\to \infty$. 
This result might indicate that there exists a terminal polydispersity, around the value 
of $0.7$, beyond which the I-N transition becomes of first order for any $\kappa_0$. We will 
certainly solve this conjecture by performing a bifurcation analysis around the tricritical point 
in such a way to find an analytic expression for 
 $\Delta_0^*$ as a function not only of $\kappa_0$ but also of higher moments of the 
distribution $f_0(\kappa_0)$. We will carry out this study in the future and we are planning to 
generalize it for any particle shape (polydisperse hard discorectangles, hard ellipses, etc.).  
In such a way we will be able to shed some light on the character (second vs. first) 
of the I-N transition that mean-field theories predict for different polydisperse particle shapes.  

\section{Conclusions}
\label{conclusions}
We have extended the SPT of HR fluid from its multicomponent version to consider a continuous length-polydispersity. 
Using this formalism we derived a simplified coexistence equations by using the Fourier expansions of the 
angular moment profiles. The numerical solution of the obtained set of equations for the Fourier components and for 
different values of the mean aspect ratio and polydispersity allowed us to obtain the phase diagram of the system.
We have not taken any functional ansatz to parameterize the density profile. Thus our calculations are exact up 
to the errors associated with the truncation of the Fourier series. 
The main results can be summarized as follows: (i) The I-N transition becomes stronger with polydispersity, 
i.e. the coexistence gap becomes wider and the tricritical point moves to higher values of $\kappa_0$. (ii) The 
stability of the T phase slightly decreases with $\Delta_0$. (iii) There exists strong fractionation between 
the coexisting phases which becomes stronger at $\kappa_0\sim 3$ and decreases as we approximate the 
T-N or I-N tricritical points. The fractionation for a Schulz-type parent distribution function 
is stronger than that obtained for a  Gaussian-tailed distribution. 
As usually the N phase is rich in long particles while the I or T phases are highly populated 
by the smaller particles. (iv) The orientational ordering of coexisting species of different $\kappa$ 
is in general different: 
the small species, those with $\kappa \gtrsim 1$, have a low orientational ordering and tend to align in T-like 
configurations while long species contribute to stabilize the uniaxial N. This effect is more pronounced 
for $\kappa_0\sim 3$. (v) The locations of tricritical points are severely affected by polydispersity and our results 
might suggest the existence of a terminal polydispersity beyond which the I-N transition becomes of first order. 
We require further calculations to definitively settle out this point.  

A recent experiment on magnetic polydisperse nanorods which are strongly confined between lamellar layers 
(an experimental realization of a quasi-2D hard-rod fluid)
shows a first order I-N transition with the N director being parallel to the plane of the layers 
\cite{Constantin}. As we show in the present work when the polydispersity is high enough the character of the I-N 
transition could change from second to first order. Thus the polydispersity could explain the experimental 
results obtained in Ref. \cite{Constantin}.   

We have not considered the presence of non-uniform phases (like smectic or crystal) in the present study. For low 
polydispersity they become stable at high densities. However when polydispersity increases they rapidly 
destabilize so the results shown here, specially those obtained for high enough $\Delta_0$, 
could be qualitatively confirmed by experimental realizations of a 2D 
hard-rectangular fluid. One of them could be mechanically vibrated monolayers of granular cylinders. 
Recent experiments on such systems have confirmed the presence of strong T, N \cite{Narayan1,Heras} 
and smectic \cite{G-P} ordering of rods, which strongly depends on the aspect ratio and packing fraction.   
Our group will conduct experiments on shaken monolayers of granular rods to study the 
effect of length-polydispersity on the stability of those liquid-crystal textures 
already found in the one-component case \cite{G-P}.  

\acknowledgements

Financial support from MINECO (Spain) under grant FIS2015-66523-P is acknowledged.

\appendix

\section{The I$_1$-N$_0$ coexistence equations}

For  $q=1$ and $\gamma_{\rm I}=1$ the integral with respect to $\kappa$ in the set of equations  
(\ref{set3}) corresponding to the I-N coexistence calculations can be performed 
analytically resulting in the following set of equations for the variables 
$\eta_j^{(i,\rm N)}$: 
\begin{eqnarray}
&&\eta_j^{(i,\rm N)}=\kappa_0^{-1}\eta_j^{(0,\rm N)}\delta_{i1}\nonumber\\&&+D_{j0}
\eta_0\left(1-\kappa_0^{-1}\right)^i\int_0^1 du 
\cos(\pi j u) 
\frac{e^{-R_0(\pi u)}}{\left[1+R_1(\pi u)\right]^{\nu+i+1}},
\nonumber\\&&
\end{eqnarray}
where we have defined the functions
\begin{eqnarray}
&&R_0(\phi)=\log\left(\frac{1-\eta_0}{1-\eta_0^{(1,N)}}\right)
\nonumber\\&&+\frac{4}{\pi}\left[\frac{\eta_0^{(1,\rm N)}+\eta_0^{(0,\rm N)}\kappa_0^{-1}-\sum_{j\geq 1} 
s_{2j}^{(\rm N)}g_{2j}\cos(2j\phi)}
{1-\eta_0^{(1,\rm N)}}\right.\nonumber\\
&&\left. -\frac{\eta_0}{1-\eta_0}(1+\kappa_0^{-1})\right],\label{R0}\\
&&R_1(\phi)=\frac{2(\kappa_0-1)}{\pi(\nu+1)}\nonumber\\&&\times\left[
\frac{\eta_0^{(1,\rm N)}+\eta_0^{(0,\rm N)}\kappa_0^{-1}-\sum_{j\geq 1} 
s_j^{(\rm N)}g_j\cos(j\phi)}
{1-\eta_0^{(1,\rm N)}}\right.\nonumber\\
&&\left. -\frac{\eta_0}{1-\eta_0}(1+\kappa_0^{-1})\right],
\end{eqnarray}
with $s_j^{(\rm N)}\equiv \eta_j^{(1,\rm N)}+(-1)^j\kappa_0^{-1}\eta_j^{(0,\rm N)}$, and 
$g_j\equiv \left(4j^2-1\right)^{-1}$. 
For the cloud I coexisting phase we have 
$\eta_0^{(i,\rm I)}=\eta_0$.

For $q=2$ and $\gamma_{\rm I}=1$  we obtain 
\begin{eqnarray}
&&\eta_j^{(i,\rm N)}=\kappa_0^{-1}\eta_j^{(0,\rm N)}\delta_{i1}+\eta_0(1-\kappa_0^{-1})^i\nonumber\\&&
\times\frac{\Gamma(\nu+i+1)}{2^{(\nu+i-1)/2}\Gamma[(\nu+i+1)/2]}
\int_0^1 du \cos(\pi j u)\nonumber\\&&\times e^{-R_0(\pi u)}e^{{\cal R}_1^2(\pi u)/8}
D_{-(\nu+i+1)}\left[{\cal R}_1(\pi u)/\sqrt{2}\right],\nonumber\\&&
\end{eqnarray}
where the function $R_0(\phi)$ is the same as (\ref{R0}), while  
\begin{eqnarray}
{\cal R}_1(\phi)=R_1(\phi)\frac{(\nu+1)\Gamma\left[(\nu+1)/2\right]}{\Gamma\left[(\nu+2)/2\right]}.
\end{eqnarray}
$D_{\mu}(x)$ is the parabolic cylinder function. We used a Fortran 
subroutine provided in Ref. \cite{parabolic} to numerically evaluate $D_{\mu}(x)$.

\end{document}